\documentclass[prb,floatfix,twocolumn,showpacs,amsmath,amssymb]{revtex4}
\usepackage{graphicx}
\usepackage{dcolumn}
\usepackage{bm}
\begin{document}

\title{Magnetism and superconductivity in the $t{-}t^\prime{-}J$ model}
\author{Leonardo Spanu, Massimo Lugas, Federico Becca, and Sandro Sorella}
\affiliation{INFM-Democritos, National Simulation Center and International
School for Advanced Studies (SISSA), I-34014 Trieste, Italy}
\date{\today}

\begin{abstract}
We present a systematic study of the phase diagram of the $t{-}t^\prime{-}J$ 
model by using the Green's function Monte Carlo (GFMC) technique, implemented 
within the fixed-node (FN) approximation and a wave function that contains both 
antiferromagnetic and d-wave pairing. This enables us to study the interplay 
between these two kinds of order and compare the GFMC results with the ones 
obtained by the simple variational approach. By using a generalization of the
forward-walking technique, we are able to calculate true FN ground-state
expectation values of the pair-pair correlation functions. In the case of 
$t^\prime=0$, there is a large region with a coexistence of superconductivity
and antiferromagnetism, that survives up to $\delta_c \sim 0.10$ for $J/t=0.2$
and $\delta_c \sim 0.13$ for $J/t=0.4$. The presence of a finite $t^\prime/t<0$
induces a strong suppression of both magnetic (with $\delta_c \lesssim 0.03$, 
for $J/t=0.2$ and $t^\prime/t=-0.2$) and pairing correlations. In particular,
the latter ones are depressed both in the low-doping regime and around 
$\delta \sim 0.25$, where strong size effects are present.
\end{abstract}

\maketitle

\section{Introduction}

After more than twenty years from the discovery of high-temperature 
superconductivity, a comprehensive description of the cuprate materials is 
still lacking. One of the main concern is about the origin of the electron 
pairing, namely if it is due to electron-phonon coupling, like in the standard 
theory by Bardeen, Cooper and Schrieffer (BCS),~\cite{bcs} or it can be 
explained by alternative mechanisms, based on the electronic interaction alone.
From one side, though the isotope effect in cuprates (if any) is much smaller 
than the one observed in BCS superconductors, there are experiments suggesting 
a strong coupling between electrons and localized lattice 
vibrations.~\cite{lanzara1,lanzara2} 
On the other side, besides a clear experimental outcome showing unusual 
behaviors in both metallic and superconducting phases, there is an increasing 
theoretical evidence that purely electronic models can indeed sustain a robust 
pairing, possibly leading to a high critical 
temperature.~\cite{sorella,maier,baeriswyl}
Within the latter scenario, the minimal microscopic model to describe the 
low-energy physics has been proposed to be the Hubbard model or its 
strong-coupling limit, namely the $t{-}J$ model, which includes 
an antiferromagnetic coupling between localized spins and a kinetic term for 
the hole motion.~\cite{anderson,rice} In this respect, Anderson proposed 
that electron pairing could naturally emerge from doping a Mott insulator, 
described by a resonating valence bond (RVB) state, where 
the spins are coupled together forming a liquid of singlets.~\cite{anderson} 
Indeed, subsequent numerical calculations for the $t{-}J$ model,~\cite{gros1} 
showed that, though the corresponding Mott insulator (described by the 
Heisenberg model) has magnetic order, the RVB wave function with d-wave 
symmetry in the electron pairing can be stabilized in a huge region of doping 
close to the half-filled insulator.
These calculations have been improved by studying the accuracy of such a 
variational state, giving solid and convincing arguments for the existence of 
a superconducting phase in the $t{-}J$ model.~\cite{sorella}
However, other numerical techniques, like Density Matrix Renormalization
Group (DMRG), provided some evidence that charge inhomogeneities can
occur at particular filling concentrations.~\cite{white1,white2} These stripes 
are probably enhanced by the strong anisotropic boundary conditions used in 
this approach and can be also found by allowing anisotropies in the hopping
and in the super-exchange coupling.~\cite{beccastripes}
Coming back to the projected RVB wave functions, it is worth mentioning that 
an approximate and simplified description of these states can be 
obtained by the renormalized mean-field theory (RMFT), the so-called 
``plain vanilla'' approach.~\cite{AZpaper} 
When this approach is applied to the $t-J$ model, it is possible to describe 
many unusual properties of the high-temperature superconductors and capture 
the most important aspects of the cuprate phase diagram.~\cite{grosrev} 
However, at present, most of the calculations have been done by neglecting 
antiferromagnetic correlations, that are definitively important at low doping. 
Within RMFT and most of the variational calculations, the magnetic correlations
are omitted, implying a spin liquid (disordered) state in the insulating 
regime. Although antiferromagnetism can be easily introduced in both 
approaches, it is often not satisfactorily described, since the presence of an 
antiferromagnetic order parameter in the fermionic determinant implies a wrong 
behavior of the spin properties at small momenta,~\cite{gros2,ivanov} unless 
a spin Jastrow factor is used to describe the corresponding spin-wave 
fluctuations. Indeed, it is now well known that the accurate correlated  
description of an ordered state is obtained by applying a long-range spin 
Jastrow factor to a state with magnetic order.~\cite{manousakis,franjio,becca} 
The important point is that the Gaussian fluctuations induced by the Jastrow 
term must be orthogonal to the direction of the order parameter, in order to 
reproduce correctly the low-energy spin-wave excitations. Moreover, by
generalizing the variational wave function to consider Pfaffians instead of 
simple determinants,~\cite{lhuillier,lugas} it is possible to consider both 
electron pairing and magnetic order, that are definitively important to 
determine the phase diagram of the $t{-}J$ model in the low-doping regime. 

The interplay between superconductivity and magnetism is the subject of an 
intense investigation in the recent years. In most of the thermodynamic 
measurements these two kinds of order do not coexist, though elastic neutron 
scattering experiments for underdoped ${\rm YBa_2Cu_3O_x}$ could suggest a 
possible coexistence, with a small staggered 
magnetization.~\cite{sidis,hodges,mook} On the contrary, in the $t{-}J$ model, 
there is an evidence in favor of a coexistence,~\cite{sorella} 
the antiferromagnetic order surviving up to a relatively large hole doping, 
i.e., $\delta \sim 0.1$ for $J/t=0.2$.~\cite{lugas} Therefore, the regime 
of magnetic order predicted by these calculations extend to much larger 
doping than the experimental results and also the robustness of the 
coexistence of superconductivity and antiferromagnetism seems to be 
inconsistent with the experimental outcome. Of course, disorder effects, which
are expected to be important especially in the underdoped region, would 
affect the general phase diagram.~\cite{dagotto} However, without invoking 
disorder, one is also interested to understand if alternative ingredients can 
modify the phase diagram of the simple $t{-}J$ model. For instance, 
band structure calculations support the presence of a sizable second-neighbor
hopping $t^\prime$ in cuprate materials, showing a possible connection between 
the value of the highest critical temperature and the ratio 
$t^\prime/t$.~\cite{pavarini} Moreover, an experimental analysis suggests an 
influence of the value of $t^\prime/t$ on the pseudogap energy 
scale.~\cite{tanaka}
From a theoretical point of view, the effect of $t^\prime$ is still not
completely elucidated,~\cite{tohyama,white3,ogata,tklee1,tklee2,oles} though 
there are different calculations providing evidence that a finite $t^\prime$ 
could suppress superconductivity in the low-doping regime.
On the other hand, recent Monte Carlo calculations suggest that the presence
of $t^\prime$ (as well as a third-neighbor hopping $t^{\prime \prime}$) could 
induce an enhancement of pairing in optimal and overdoped 
regions.~\cite{tklee1,tklee2}

In this paper, we want to examine the problem of the interplay between
magnetism and superconductivity in the $t{-}J$ model and its extension 
including a next-nearest-neighbor hopping $t^\prime$ by using
improved variational and Green's function Monte Carlo (GFMC) techniques.
Indeed, especially the latter approach has been demonstrated to be very 
efficient in projecting out a very accurate approximation of the exact
ground state and, therefore, can give useful insight into this
important issue related to high-temperature superconductivity.

The paper is organized as follows: In Sec.~\ref{model} we describe the methods
we used, in Sec.~\ref{results} we show the numerical results for 
antiferromagnetism and superconductivity, and in Sec.~\ref{conc} we draw our
conclusions.

\section{Model and Method}\label{model}

\subsection{Model and variational wave function}

We consider the $t{-}t^\prime{-}J$ model on a two-dimensional square lattice
with $L$ sites and periodic boundary conditions on both directions:
\begin{eqnarray}\label{hamitj}
{\cal H} &=& J \sum_{\langle i,j \rangle} \left ( {\bf S}_i \cdot {\bf S}_j - 
\frac{1}{4} n_i n_j \right ) \nonumber \\
&-& t \sum_{\langle i,j \rangle \sigma} c_{i,\sigma}^{\dagger} c_{j,\sigma} 
- t^\prime \sum_{\langle \langle k,l \rangle \rangle \sigma}
c_{k,\sigma}^{\dagger} c_{l,\sigma} + h.c.
\end{eqnarray}
where $\langle \dots \rangle$ indicates the nearest-neighbor sites, 
$\langle \langle \dots \rangle \rangle$ the next-nearest-neighbor sites,
$c_{i,\sigma}^{\dagger}$ ($c_{i,\sigma}$) creates (destroys) an electron 
with spin $\sigma$ on the site $i$, ${\bf S}_i =(S^x_i,S^y_i,S^z_i)$ is the 
spin operator, $S^\alpha_i = \sum_{\sigma,\sigma^\prime} c_{i,\sigma}^{\dagger} 
\tau^\alpha_{\sigma,\sigma^\prime} c_{i,\sigma^\prime}$, being $\tau^\alpha$ 
the Pauli matrices, and 
$n_i = \sum_{\sigma} c_{i,\sigma}^{\dagger} c_{i,\sigma}$ is the local density 
operator. In the following, we set $t=1$ and consider $t^\prime=0$ and 
$t^\prime/t=-0.2$. Moreover, we consider two kinds of square clusters:
Standard clusters with $L=l \times l$ sites and $45^\circ$ tilted lattices with 
$L=2 \times l^2$ sites. Besides translational symmetries, both of them 
have all reflection and rotational symmetries.

The variational wave function is defined by:
\begin{equation}\label{wf}
|\Psi_{VMC} \rangle = 
{\cal J}_s {\cal J}_d {\cal P}_N {\cal P}_G |\Phi_{MF}\rangle,
\end{equation}
where ${\cal P}_G$ is the Gutzwiller projector that forbids double occupied 
sites, ${\cal P}_N$ is the projector onto the subspace with fixed number of 
$N$ particles. Moreover, ${\cal J}_s$ is a spin Jastrow factor 
\begin{equation}\label{spinjastrow}
{\cal J}_s = \exp \left ( \frac{1}{2} \sum_{i,j} v_{ij} S^z_i S^z_j \right ),
\end{equation}
being $v_{ij}$ variational parameters, and finally ${\cal J}_d$ is a
density Jastrow factor
\begin{equation}\label{densityjastrow}
{\cal J}_d = \exp \left ( \frac{1}{2} \sum_{i,j} u_{ij} n_i n_j \right ),
\end{equation} 
being $u_{ij}$ other variational parameters. 
The above wave function can be efficiently sampled by standard variational
Monte Carlo, by employing a random walk of a configuration $|x \rangle$,
defined by the electron positions and their spin components along the $z$
quantization axis. Indeed, in this case, both Jastrow terms are very simple 
to compute, since they only represent classical weights acting on the
configuration.

As previously reported,~\cite{lugas} the main difference from previous 
approaches is the presence of the spin Jastrow factor and the choice 
of the mean-field state $|\Phi_{MF} \rangle$, defined as the ground state of
the mean-field Hamiltonian
\begin{eqnarray}
\label{meanfield}
&& {\cal H}_{MF} =  
\sum_{i,j,\sigma} t_{i,j} c_{i,\sigma}^{\dagger} c_{j,\sigma} + h.c. 
-\mu \sum_{i,\sigma} n_{i,\sigma}  \nonumber \\
&&+ \sum_{\langle i,j \rangle} \Delta_{i,j} 
(c_{i,\uparrow}^{\dagger} c_{j,\downarrow}^{\dagger} + 
c_{j,\uparrow}^{\dagger} c_{i,\downarrow}^{\dagger} + h.c.) + {\cal H}_{AF},
\end{eqnarray}
where we include both BCS pairing $\Delta_{i,j}$ [with $d$-wave symmetry,
i.e., for nearest-neighbor sites $\Delta_k = \Delta (\cos k_x - \cos k_y)$] 
and staggered magnetic field in the $x{-}y$ plane 
\begin{equation}\label{hamaf}
{\cal H}_{AF} = \Delta_{AF} \sum_i (-1)^{R_i}
(c_{i,\uparrow}^{\dagger}c_{i,\downarrow} + 
c_{i,\downarrow}^{\dagger}c_{i,\uparrow}), 
\end{equation}
where $\Delta_{AF}$ is a variational parameter that, together with the
chemical potential $\mu$ and the next-nearest-neighbor hopping of
Eq.~(\ref{meanfield}), can be determined by minimizing the variational energy
of ${\cal H}$. Whenever both $\Delta$ and $\Delta_{AF}$ are finite, the
projection $\langle x|\Phi_{MF} \rangle$ of the mean-field state on a given 
configuration $|x \rangle$ can be described in terms of Pfaffians,
instead if $\Delta=0$ or $\Delta_{AF}=0$ it can be described by using
determinants. Moreover, only in the case where the magnetic order parameter 
is in the $x{-}y$ plane, the presence of the spin Jastrow 
factor~(\ref{spinjastrow}) can introduce relevant fluctuations over the 
mean-field order parameter $\Delta_{AF}$, leading to an accurate description 
of the spin properties. A detailed description of the wave function of
Eq.~(\ref{wf}) and its physical properties can be found in 
Ref.~\onlinecite{lugas}. The variational parameters contained in the mean-field
Hamiltonian~(\ref{meanfield}) and in the Jastrow factors~(\ref{spinjastrow}) 
and~(\ref{densityjastrow}) are calculated by using the optimization technique 
described in Ref.~\onlinecite{sorella2}, that makes it possible to handle with 
a rather large number of variational parameters.

\subsection{GFMC: beyond the Variational Monte Carlo}

The optimized variational wave function $|\Psi_{VMC} \rangle$ can be also used 
within the GFMC method to filter out an approximation of the ground state
$|\Psi_0^{FN} \rangle$. Indeed, due to the presence of the fermionic sign 
problem, in order to have a stable numerical calculation, the GFMC must be 
implemented within the fixed-node (FN) approach, that imposes to 
$|\Psi_0^{FN} \rangle$ to have the same nodal structure of the 
variational ansatz.~\cite{ceperley} Here, we recall the basic definitions of 
the standard FN method. A detailed description of this technique can be found 
in Ref.~\onlinecite{lugas}. 

Starting from the original Hamiltonian ${\cal H}$, we define an effective 
model by
\begin{equation}\label{fnham}
{\cal H}_{eff} = {\cal H} + O.
\end{equation}
The operator $O$ is defined through its matrix elements and depends upon
a given {\it guiding} function $|\Psi \rangle$, that is for instance the 
variational state itself, i.e., $|\Psi_{VMC} \rangle$:
\begin{equation} \label{defo}
O_{x^\prime,x} = \left \{
\begin{array}{ll}
-{\cal H}_{x^\prime,x} & {\rm if} \; s_{x^\prime,x} = \Psi_{x^\prime} {\cal H}_{x^\prime,x} \Psi_x >0 \nonumber \\
\sum_{y,s_{y,x}>0} {\cal H}_{y,x} \frac{\Psi_y}{\Psi_x} & {\rm for} \; x^\prime=x,
\end{array}
\right .
\end{equation}
where $\Psi_x = \langle x|\Psi \rangle$, $|x \rangle$ being an electron
configuration with definite $z$-component of the spin.
Notice that the above operator annihilates the guiding function, namely
$O |\Psi \rangle=0$. Therefore, whenever the guiding function is close
to the exact ground state of ${\cal H}$, the perturbation $O$ is expected to 
be small and the effective Hamiltonian becomes very close to the original one.
The most important property of this effective Hamiltonian is that its ground 
state $|\Psi_0^{FN} \rangle$ can be efficiently computed by using 
GFMC.~\cite{nandini,calandra} The distribution
$\Pi_x \propto \langle x|\Psi \rangle \langle x|\Psi_0^{FN} \rangle$
is sampled by means of a statistical implementation of the power method:
$\Pi \propto \lim_{n \to \infty } G^n \Pi^0$, where
$\Pi^0$ is a starting distribution and $G_{x^\prime,x}= \Psi_{x^\prime}
(\Lambda \delta_{x^\prime,x} - {\cal H}_{eff,x^\prime,x})/ \Psi_x$,
is the so-called Green's function, $\delta_{x^\prime,x}$ being the Kronecker 
symbol. The statistical method is very efficient since all the matrix 
elements of $G$ are non-negative and, therefore, $G$ can represent a transition
probability in configuration space, apart for a normalization factor
$b_x= \sum_{x^\prime} G_{x^\prime,x}$.
Since $|\Psi_0^{FN} \rangle$ is an exact eigenstate of the effective 
Hamiltonian ${\cal H}_{eff}$, the corresponding ground-state energy can be 
evaluated efficiently by computing  
\begin{equation}\label{enma}
E_{MA}= \frac{\langle \Psi_{VMC} | {\cal H}_{eff} | \Psi_0^{FN} \rangle}
{\langle \Psi_{VMC}|\Psi_0^{FN} \rangle},
\end{equation}
namely the statistical average of the local energy 
$e_L(x)=\langle \Psi_{VMC}|{\cal H}|x \rangle / \langle \Psi_{VMC}|x \rangle$ 
over the distribution $\Pi_x$. The quantity $E_{MA} \le E_{VMC}$ because, by 
the variational principle 
$E_{MA} \le \langle \Psi|{\cal H}_{eff}|\Psi \rangle / \langle 
\Psi|\Psi \rangle = E_{VMC}$.
Moreover, $E_{MA}$ represents an upper bound of the expectation value $E_{FN}$ 
of ${\cal H}$ over $|\Psi_0^{FN} \rangle$, as it is shown in 
Ref.~\onlinecite{ceperley} or it can be simply derived by considering that 
the operator $O$ is semi-positive definite, namely all its eigenvalues are 
non-negative. In the following, we will denote by FN the (variational)
results obtained by using the GFMC method with fixed-node approximation,
whereas the standard variational Monte Carlo results obtained by considering
the wave function of Eq.~(\ref{wf}) will be denoted by VMC.

Summarizing the FN approach is a more general and powerful variational method
than the straightforward variational Monte Carlo. Within the FN method,
the wave function $|\Psi_0^{FN} \rangle$, the ground state of ${\cal H}^{eff}$ 
is known only statistically, and, just as in the variational approach, 
$E_{FN}$ depends explicitly on the variational parameters defining the
guiding function $|\Psi \rangle$. This is due to the fact that ${\cal H}_{eff}$
depends upon $|\Psi \rangle$ through the operator $O$. The main advantage of 
the FN approach is that it provides the exact ground-state wave function for 
the undoped insulator (where the signs of the exact ground state are known), 
and therefore it is expected to be particularly accurate in the important 
low-doping region. Moreover, the FN method is known to be very efficient in
various cases: For instance, it has allowed to obtain a basically exact
description of the three-dimensional system of electrons interacting through
the realistic Coulomb potential (in presence of a uniform positive 
background).~\cite{alder} Therefore, it represents a very powerful tool to
describe correlated electronic systems. 

\section{Results}\label{results}

\subsection{Phase separation}

Before showing the results on magnetic and superconducting properties, we
briefly discuss the stability against phase separation. 
In order to detect a possible phase separation, it is very useful to follow
the criterion given in Ref.~\onlinecite{lin} and consider the energy per hole:
\begin{equation}\label{eh}
e_h(\delta)=\frac{e(\delta)-e(0)}{\delta},
\end{equation}
where $e(\delta)$ is the energy per site at hole doping $\delta$ and $e(0)$ is
its value at half filling. In practice, $e_h(\delta)$ represents the chord 
joining the energy per site at half filling and the one at doping $\delta$. 
For a stable system, $e_h(\delta)$ must be a monotonically increasing function 
of $\delta$, because the ground-state energy of a short-range Hamiltonian 
is a convex function of the doping. By performing exact energy calculations 
on finite clusters, the phase separation instability is marked by a minimum of 
$e_h(\delta)$ at a given $\delta_c$ and by a flat behavior up to $\delta_c$ in 
the thermodynamic limit. For an approximate variational technique based on a 
spatially homogeneous ansatz, this flat behavior is never reached in the phase 
separated region, so that $e_h(\delta)$ remains with a well defined minimum 
even for very large sizes; in this case, $\delta_c$ can be estimated with the 
Maxwell construction, provided the variational asnatz is accurate enough.
In Ref.~\onlinecite{lugas}, we demonstrated the existence of an homogeneous 
state for $t^\prime=0$ and $J/t \lesssim 0.7$. As shown in Table~\ref{tab1},
the FN approximation, that is exact at zero doping,~\cite{lugas} 
provides a substantial lowering of the VMC energy, especially away from
half filling and for a finite $t^\prime$. This is a first indication that, 
for $t^\prime/t < 0$, the simple variational approach could not be adequate 
to provide a reliable quantitative description of the ground-state properties. 

\begin{figure}
\includegraphics[width=0.45\textwidth,clip]{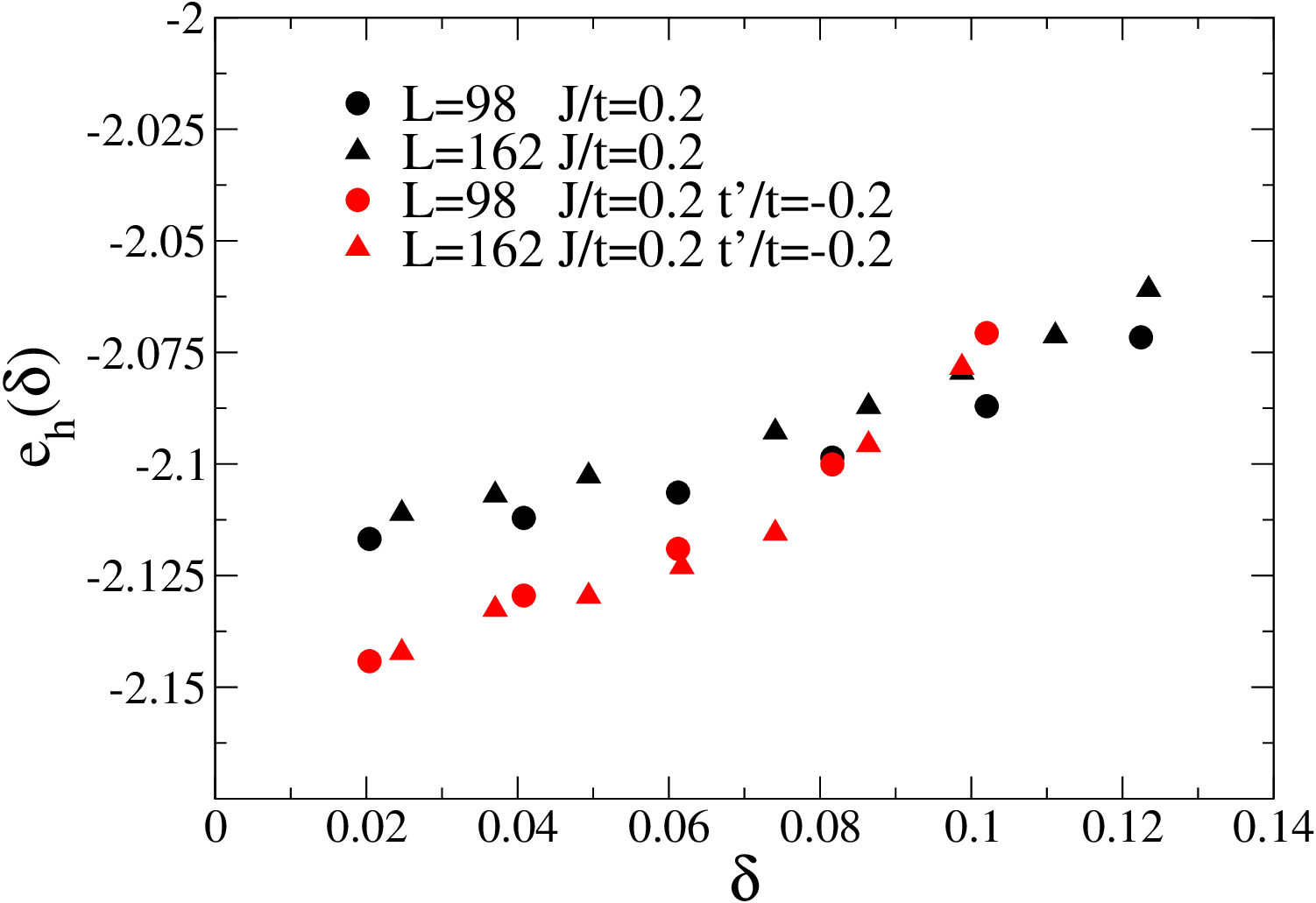}
\caption{\label{fig:emery}
(Color online) Energy per hole $e_h(\delta)$, calculated by using the FN 
method, as a function of the doping $\delta$ for $L=98$ and $162$ and two 
values of the next-nearest-neighbor hopping $t^\prime/t=0$ and $-0.2$.}
\end{figure}

The FN results clearly indicate that the inclusion of a negative 
next-nearest-neighbor hopping contributes to further stabilize the homogeneous 
phase at finite doping, see Fig.~\ref{fig:emery}.
This result is compatible with the outcome of recent calculations based on 
cluster dynamical mean-field theory on the Hubbard model, where a negative 
ratio $t^\prime/t$ enhances the stability of the homogeneous phase, whereas 
positive values of $t^\prime$ favor phase separation.~\cite{macridin}
In this work, we do not want to address in much detail this issue and we
will focus our attention on the more interesting magnetic and superconducting
properties.

\begin{table}
\caption{\label{tab1} 
Variational (VMC) and fixed-node (FN) energies per site for $J/t=0.2$ and 
$t^\prime=0$ (third and fourth columns), and $t^\prime/t=-0.2$ (fifth and 
sixth columns) for two clusters with $L=98$ and $162$ and different hole
concentrations $N_h=L-N$}
\begin{tabular}{cccccc}
\hline
$L$ & $N_h$ & $E_{VMC}/L$  & $E_{FN}/L$  & $E_{VMC}/L$  & $E_{FN}/L$  \\
\hline \hline
98  &  0    & -0.233879(1) & -0.23432(1) & -0.233879(1) & -0.23432(1) \\
98  &  2    & -0.274144(5) & -0.27752(1) & -0.27290(1)  & -0.27808(1) \\
98  &  4    & -0.31429(1)  & -0.32053(1) & -0.31189(1)  & -0.32123(1) \\
98  &  6    & -0.35482(1)  & -0.36328(1) & -0.35132(1)  & -0.36405(1) \\
98  &  8    & -0.39550(1)  & -0.40563(2) & -0.39028(1)  & -0.40575(1) \\
98  & 10    & -0.43581(1)  & -0.44728(2) & -0.42814(1)  & -0.44561(1) \\
162 &  0    & -0.233707(1) & -0.23409(1) & -0.233707(1) & -0.23409(1) \\
162 &  2    & -0.258002(5) & -0.26020(1) & -0.257260(5) & -0.26012(1) \\
162 &  4    & -0.282117(5) & -0.28621(1) & -0.28067(1)  & -0.28698(1) \\
162 &  6    & -0.306324(5) & -0.31212(1) & -0.30429(1)  & -0.31307(1) \\
162 &  8    & -0.33060(1)  & -0.33793(1) & -0.32807(1)  & -0.33925(2) \\
162 & 10    & -0.35498(1)  & -0.36360(2) & -0.35207(1)  & -0.36514(2) \\
162 & 12    & -0.37954(1)  & -0.38912(2) & -0.37567(1)  & -0.39079(2) \\
162 & 14    & -0.40406(1)  & -0.41446(2) & -0.39939(1)  & -0.41520(2) \\
162 & 16    & -0.42838(1)  & -0.43946(2) & -0.42232(1)  & -0.43936(2) \\
\hline \hline
\end{tabular}
\end{table}

\subsection{Antiferromagnetic properties}

Here we present the results for the magnetic properties of the 
$t{-}t^\prime{-}J$ model and compare the FN approach with the VMC one, based 
upon the wave function~(\ref{wf}). 
As already discussed in Ref.~\onlinecite{lugas}, the optimized wave 
function~(\ref{wf}) breaks the SU(2) spin symmetry, because of the magnetic 
order parameter $\Delta_{AF}$ of Eq.~(\ref{hamaf}) and the spin Jastrow 
factor~(\ref{spinjastrow}). It turns out that at half-filling and in the 
low-doping regime, the variational state~(\ref{wf}) has an antiferromagnetic 
order in the $x{-}y$ plane, whereas the spin-spin correlations in the $z$ axis
decay very rapidly. Therefore, in order to assess the magnetic order at the 
variational level, we have to consider the isotropic spin-spin correlations:
\begin{equation}
\langle {\bf S}_0 \cdot {\bf S}_r \rangle
= \frac{\langle \Psi_{VMC}|{\bf S}_0 \cdot {\bf S}_r | \Psi_{VMC} \rangle}
{\langle \Psi_{VMC}|\Psi_{VMC} \rangle}.
\end{equation}
The FN approach alleviates the anisotropy between the $x{-}y$ plane and the 
$z$ axis; in this case, we find that a rather accurate (and much less 
computational expensive) way to estimate the magnetic moment can be
obtained from the $z$ component of the spin-spin correlations: 
\begin{equation}
\langle S_0^z S_r^z \rangle
= \frac{\langle \Psi_0^{FN}|S_0^z S_r^z | \Psi_0^{FN} \rangle}
{\langle \Psi_0^{FN}|\Psi_0^{FN} \rangle}.
\end{equation}
This quantity can be easily computed within the forward-walking 
technique,~\cite{calandra} because the operator $S_0^z S_r^z$ is diagonal in 
the basis of configurations used in the Monte Carlo sampling.
From the spin-spin correlations at the maximum distance, it is possible to
extract the value of the magnetization. In particular, the variational 
wave function is not a singlet when the antiferromagnetic order sets in,
and the magnetization has to be computed with the spin isotropic expression 
$M=\lim_{r \to \infty} \sqrt{\langle {\bf S}_0 \cdot {\bf S}_r \rangle}$.
On the other hand, the FN ground state is almost a perfect singlet for all 
the cases studied and the magnetization can be estimated more efficiently by
$M=\lim_{r \to \infty} \sqrt{3 \langle S_0^z S_r^z \rangle}$. 
The spin isotropy of the FN wave function can be explicitly checked by 
computing the mixed-average of the total spin square
\begin{equation}\label{s2ma}
\langle S^2 \rangle_{MA}
= \frac{\langle \Psi_{VMC}|S^2 | \Psi_0^{FN} \rangle}
{\langle \Psi_{VMC}|\Psi_0^{FN} \rangle},
\end{equation}
that vanishes if $|\Psi_0^{FN} \rangle$ is a perfect singlet, even when 
$|\Psi_{VMC} \rangle$ has not a well-defined spin value.

\begin{figure}
\includegraphics[width=0.45\textwidth,clip]{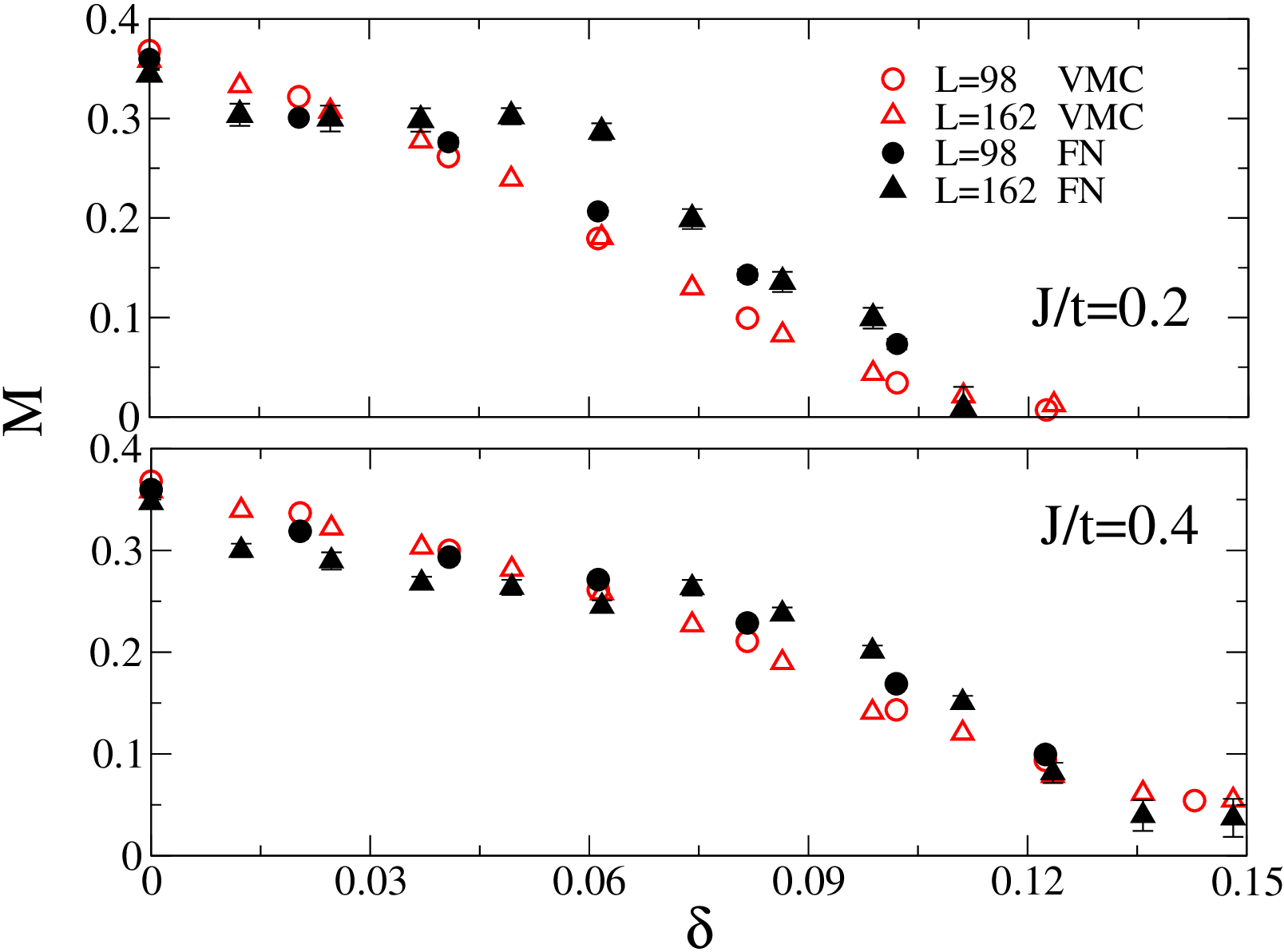}
\caption{\label{fig:AF}
(Color online) Magnetization obtained from the spin-spin correlations at the 
maximum distance calculated for the $t{-}J$ model with $J/t=0.2$ (upper panel) 
and $J/t=0.4$ (lower panel). For the VMC calculations the error-bars are 
smaller than the symbol sizes. The VMC magnetization has been obtained from 
the isotropic correlations, whereas the FN one from the correlations along the
$z$ axis (see text).}
\end{figure}

\begin{figure}
\includegraphics[width=0.45\textwidth,clip]{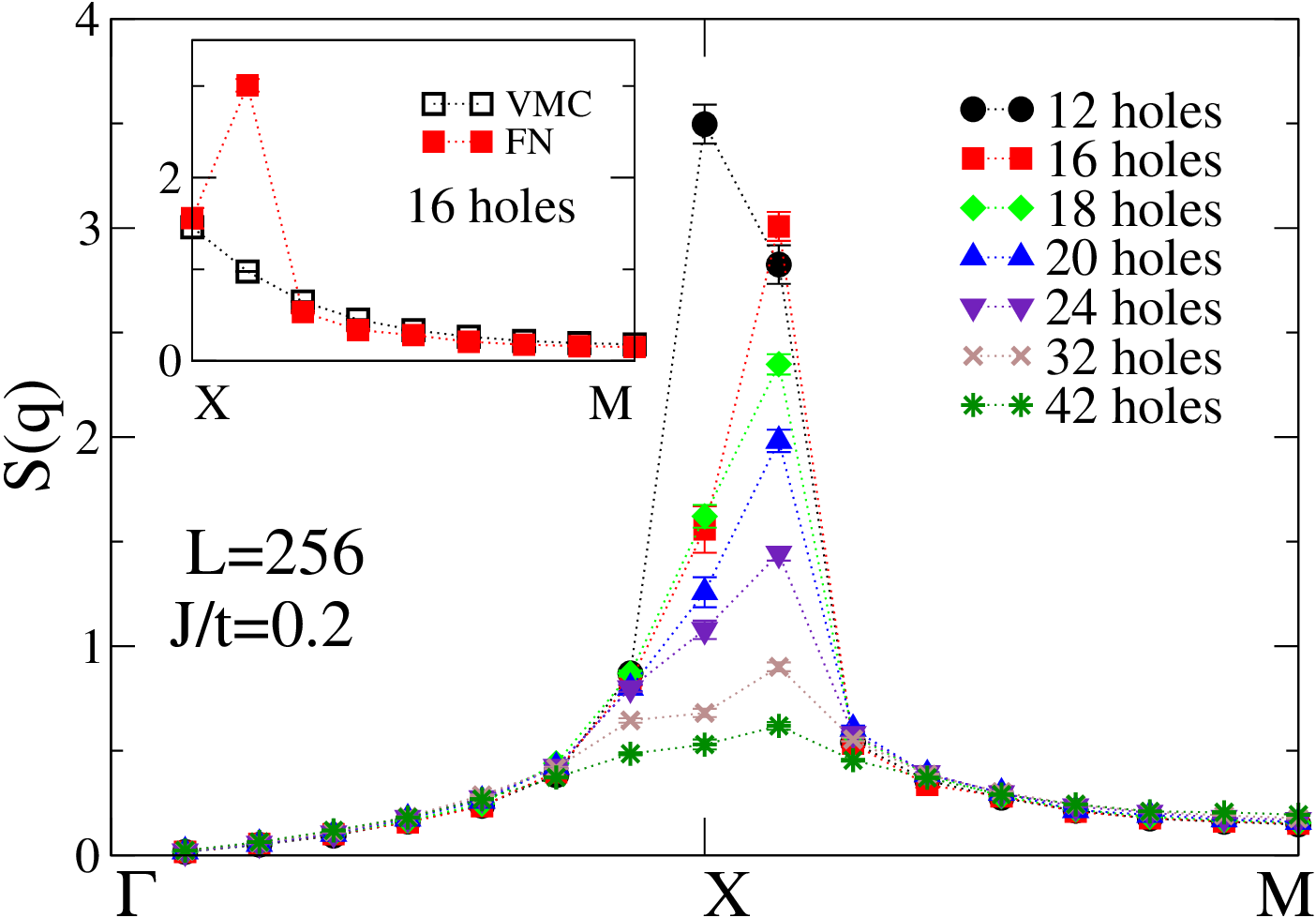}
\caption{\label{fig:SAF}
(Color online) Static spin structure factor $S(q)$ for $L=16 \times 16$ cluster
and different hole concentrations for the $t{-}J$ model with $J/t=0.2$. 
$\Gamma=(0,0)$, $X=(\pi,\pi)$, and $M=(\pi,0)$. Inset: $S(q)$ for the 
variational state (empty symbols) and for the FN approximation (full symbols).}
\end{figure}

\begin{figure}
\includegraphics[width=0.45\textwidth,clip]{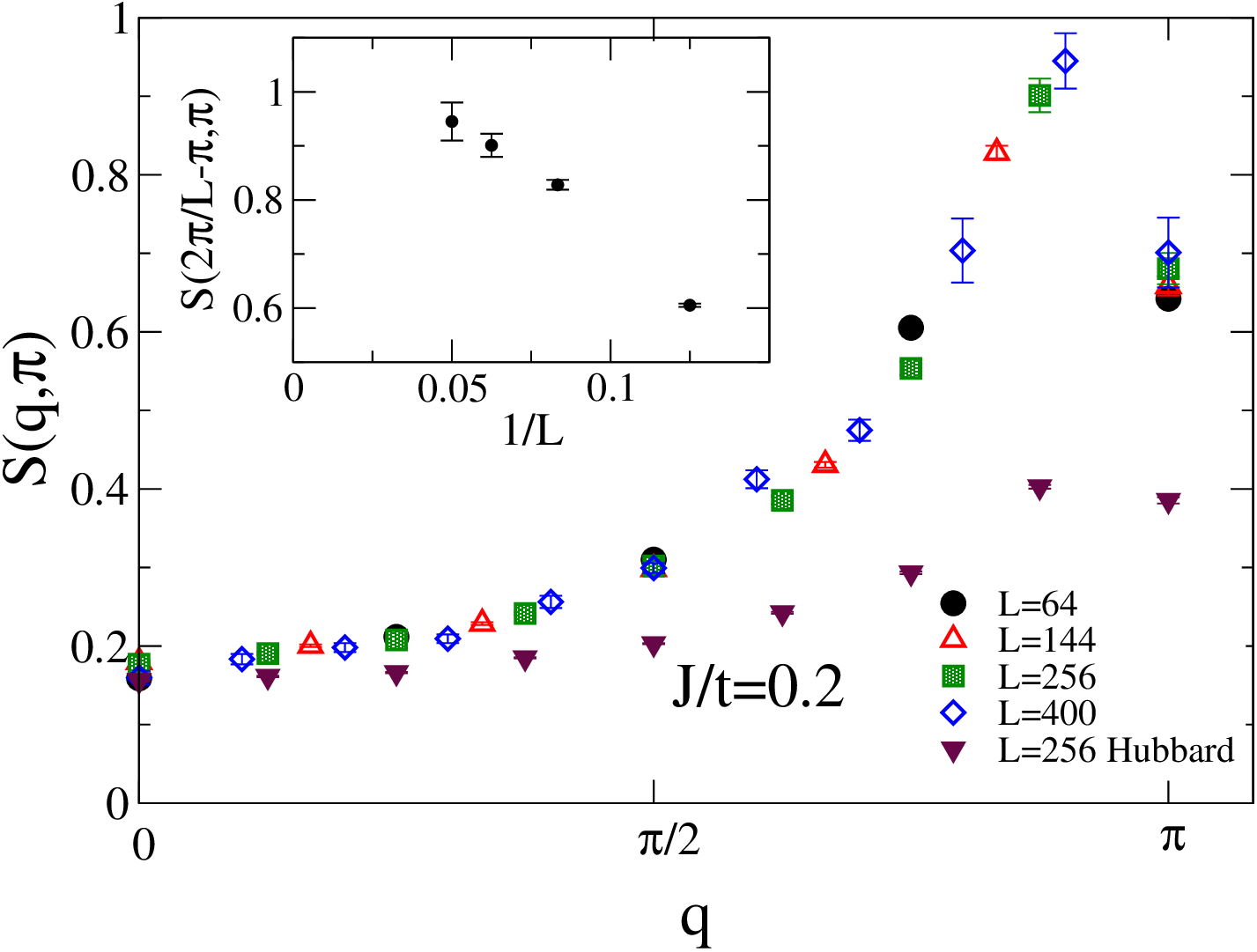}
\caption{\label{fig:scal}
(Color online) Spin structure factor $S(q)$ for the $t{-}J$ with $J/t=0.2$ and 
doping $\delta=1/8$ and different clusters ($L=8 \times 8$, $12 \times 12$, 
$16 \times 16$, and $20 \times 20$). The case of the Hubbard model at $U/t=4$
and $L=16 \times 16$ is also reported for comparison. 
Inset: Size scaling of the peak as a function of $1/L$.}
\end{figure}

In Fig.~\ref{fig:AF} we report the results of the magnetization in the
$t{-}J$ model with $J/t=0.2$ and $0.4$. 
At finite doping, it is not possible to perform a precise size scaling 
extrapolation since it is very rare to obtain the same doping concentration 
for different cluster sizes. Moreover, though the FN approach is able to 
recover an exact singlet state at half filling, $\langle S^2 \rangle_{MA}$
increases by doping, reaching its maximum around $\delta \sim 0.06$,
e.g., $\langle S^2 \rangle_{MA} \sim 1$ for $8$ holes on $162$ sites. This 
could explain why the FN results are a bit larger than the VMC ones for 
$\delta \sim 0.06$, especially for $J/t=0.2$. Definitively, both the VMC and 
FN wave functions are almost spin singlets close to the transition point, 
because the mean-field order parameter $\Delta_{AF}$ goes to zero together 
with the parameters defining the spin Jastrow factor. Therefore, we are rather 
confident in the estimation of the critical doping $\delta_c$, where the 
long-range antiferromagnetic order disappears. In particular, we find 
$\delta_c=0.10 \pm 0.01$ and $\delta_c=0.13 \pm 0.02$ for $J/t=0.2$ and 
$J/t=0.4$, respectively.

At low doping, we have evidence that long-range order is always commensurate, 
with a (diverging) peak at $X=(\pi,\pi)$ in the static spin structure factor, 
defined as
\begin{equation}
S(q)=\frac{1}{L}\sum_{l,m} e^{iq(R_l-R_m)} S_l^z S_m^z.
\end{equation}
This outcome is clear for all kinds of cluster considered, namely both for
standard $l \times l$ and $45^\circ$ tilted lattices.
By contrast, close to the critical doping $\delta_c$, we have the indication 
that some incommensurate peaks develop. Remarkably, we do not find any strong 
doping dependence of the peak positions. We show the results of $S(q)$ for the 
$16 \times 16$ cluster and $J/t=0.2$ in Fig.~\ref{fig:SAF}, where the evolution
of the peak as a function of the doping $\delta$ is reported. By increasing 
the hole doping, the commensurate peak at $X$ reduces its intensity and 
eventually the maximum of $S(q)$ shifts to a different k-point, 
i.e., $(\pi,\pi-2\pi/L)$. It should be stressed that this outcome is obtained 
only when the FN projection is applied to the lowest-energy ansatz contaning 
a sizable BCS parameter, and the FN calculation with a fully projected 
free-electron determinant cannot reproduce an incommensurate peak in $S(q)$. 
Moreover, the variational wave functions always show commensurate correlations,
see inset of Fig.~\ref{fig:SAF}.
The strong dependence of this feature on the variational ansatz may also 
indicate that more accurate calculations are necessary to clarify this 
important aspect of the phase diagram of the $t{-}J$ model. In order to 
support the validity and the accuracy of our results, we have applied the same 
method to the Hubbard model at $U/t=4$, where essentially exact calculations 
are available for $S(q)$.~\cite{imada} In this case, we have reproduced both 
the position and the intensity of the incommensurate peak on the $10 \times 10$
lattice. It is interesting to notice that, within the FN approximation, 
the intensity of the incommensurate peak at $U/t=4$ is much smaller than the 
corresponding one for the $t{-}J$ model, see Fig.~\ref{fig:scal}. 
This clearly indicates that in the $t{-}J$ model the magnetic correlations are 
much more pronounced than the corresponding ones of the Hubbard model,
possibly explaining the origin of the large extention of the antiferromagnetic 
region found in the $t{-}J$ model.

We now discuss whether these incommensurate spin correlations remain in the 
thermodynamic limit. For all the cluster sizes we considered, i.e., up to 
$L= 20 \times 20$, the incommensurate peak in $S(q)$ always appears at 
$(\pi,\pi-2\pi/L)$, namely the closest k-point to $X$ along the border of the 
Brillouin zone. This indicates that, in the thermodynamic limit, the peak 
should be located very close to $X$ and it is not compatible with 
$(\pi,\pi-2 \pi \delta)$, found in cuprate materials.~\cite{yamada}
Although size scaling extrapolations are not possible for a generic hole 
doping, we do not have evidence that the incommensurate peak diverges, implying
no incommensurate long-range order at finite doping concentrations. 
Nevertheless, once the commensurate magnetic order is melted, the ground state 
is characterized by short-range incommensurate spin correlations. 
In Fig.~\ref{fig:scal}, we show the results for $J/t=0.2$ and $\delta=1/8$, 
where different clusters with the same doping are available. Similar 
calculations with $t^\prime/t=-0.2$ show the same qualitative behavior for 
the $S(q)$.

\begin{figure}
\includegraphics[width=0.45\textwidth,clip]{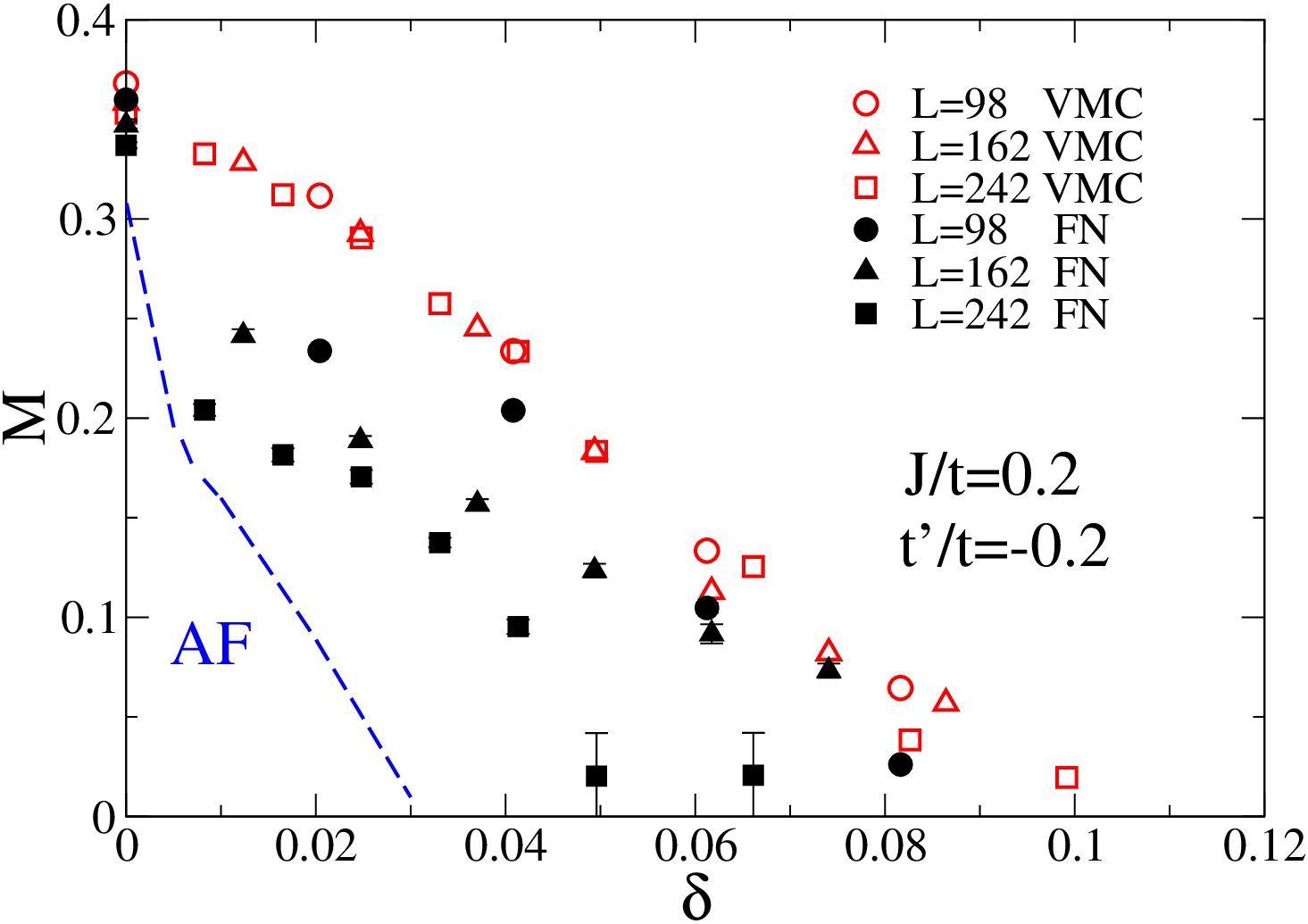}
\caption{\label{fig:AFtprime}
(Color online) The same as in Fig.~\ref{fig:AF} for the $t{-}t^\prime{-}J$ 
model with $J/t=0.2$ and $t^\prime/t=-0.2$. For the VMC calculations the 
error-bars are smaller than the symbol sizes. The dashed line indicates a
tentative estimation for the thermodynamic limit.}
\end{figure}

Coming back to the commensurate magnetic order close to half-filling, we
stress that the pure $t{-}J$ model shows robust antiferromagnetic correlations,
with a critical doping much larger than the one observed in the hole-doped 
cuprates materials, where the long-range order disappears at 
$\delta_c \sim 0.05$.~\cite{yamada} This smaller value of the critical doping 
cannot be explained by reducing the antiferromagnetic super-exchange $J$, 
given the fact that even for $J/t=0.2$ we have that $\delta_c \sim 0.1$. 
Besides disorder effects that can be important in the underdoped 
regime,~\cite{dagotto} one important ingredient to be considered in a 
microscopic model is the next-nearest-neighbor hopping, that was shown to have 
remarkable effects on both magnetic and superconducting 
properties.~\cite{tohyama,white3,tklee1,tklee2}
In particular, in spite exact diagonalization calculations suggest a 
suppression of antiferromagnetic correlations for negative 
$t^\prime/t$,~\cite{tohyama} more recent Monte Carlo simulations (also 
including a further third-neighbor hopping $t^{\prime \prime}$) do not confirm
these results, pointing instead toward a suppression of superconducting 
correlations.~\cite{tklee2}

In Fig.~\ref{fig:AFtprime}, we report the magnetization for $J/t=0.2$ 
and $t^\prime/t=-0.2$. The first outcome is that the VMC results, though 
renormalized with respect to the case $t^\prime=0$, present a critical
doping $\delta_c$ which is very similar to the one found for the pure $t{-}J$ 
model. By contrast, the FN approach strongly suppresses the spin-spin 
correlations, even very close to half filling. In this case, the FN results 
have rather large size effects, that prevent us to extract a reliable estimate 
for the thermodynamic limit. However, it is clear that the antiferromagnetic 
region is tiny and we can estimate that $\delta_c \lesssim 0.03$. It should be 
emphasized that for $t^\prime/t=-0.2$ the variational wave function is not
as accurate as for the pure $t{-}J$ model with $t^\prime=0$, but nevertheless 
the projection technique, even if approximate, is able to reduce the bias
(e.g., the presence of a large magnetic order up to $\delta \sim 0.1$), 
showing the importance of alternative numerical methods to assess the 
actual accuracy of the simple variational approach. Indeed, we are confident
that our FN results represent a good approximation of the true ground-state 
properties. On the contrary, the VMC calculations clearly show that the wave 
function~(\ref{wf}) overestimates the correct value of the magnetic moment.

\subsection{Superconducting properties}

In the following, we want to address the problem of the superconducting 
properties of the Hamiltonian~(\ref{hamitj}). In particular, we would like to 
obtain an accurate determination of the pair-pair correlations as a function 
of the hole doping and clarify the role of the next-nearest-neighbor hopping 
$t^\prime$. The effect of such term has been recently considered by using 
different numerical techniques. DMRG calculations for 
$n$-leg ladders (with $n=4$ and $6$) showed that the effect of a negative 
$t^\prime$ is to stabilize a metallic phase, without superconducting 
correlations.~\cite{white3} Moreover, improved variational 
Monte Carlo techniques suggested that $t^\prime$ could suppress pairing
at low doping, whereas some increasing of superconducting correlations can
be found in the optimal doping regime.~\cite{tklee1,tklee2} 
A further variational study,~\cite{ogata} suggested the possibility that
a sufficiently large ratio $t^\prime/t$ can disfavor superconductivity and
stabilize charge instabilities (stripes) near $1/8$ doping.

The pair-pair correlations are defined as 
\begin{equation}
\Delta^{\mu,\nu}(r)= S_{r,\mu}S^{\dagger}_{0,\nu},
\end{equation}
where $S^{\dagger}_{r,\nu}$ creates a singlet pair of electrons in the
neighboring sites $r$ and $r+\mu$, namely
\begin{equation}
S_{r,\mu}^{\dagger}=c^{\dagger}_{r,\uparrow}c^{\dagger}_{r+\mu,\downarrow}
-c^{\dagger}_{r,\downarrow}c^{\dagger}_{r+\mu,\uparrow}.
\end{equation}
For the first time, we implemented the forward-walking technique in order to 
compute true expectation values of the pairing correlations over the FN state:
\begin{equation}\label{pair}
\langle \Delta^{\mu,\nu}(r) \rangle
=\frac{\langle \Psi_0^{FN}|\Delta^{\mu,\nu}(r)| \Psi_0^{FN} \rangle}
{\langle \Psi_0^{FN}|\Psi_0^{FN} \rangle}.
\end{equation}
Indeed, given the fact that $\Delta^{\mu,\nu}(r)$ is a non-diagonal operator 
(in the basis of configurations defined before), within the FN approach the 
previous calculations~\cite{sorella} were based upon the so-called mixed 
average, where, similarly to Eqs.~(\ref{enma}) and~(\ref{s2ma}), the state on 
the left is replaced by the variational one. Now, by using Eq.~(\ref{pair}), 
it is possible to verify the fairness of the variational results against a 
much more accurate estimation of the exact correlation functions given by the 
FN approach.

\begin{figure}
\includegraphics[width=0.45\textwidth,clip]{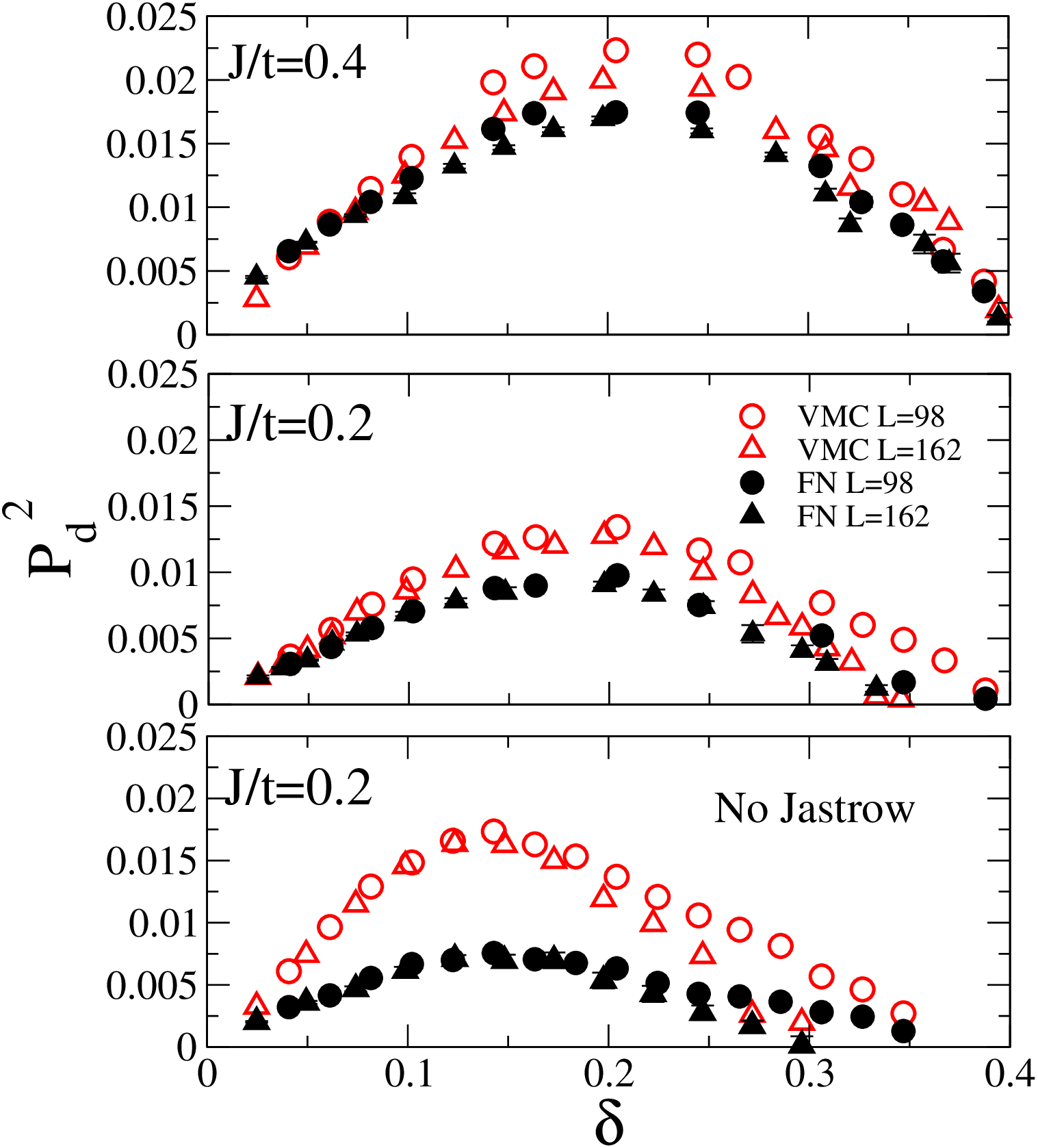}
\caption{\label{fig:pair}
(Color online) Pair-pair correlations at the maximum distance as a function on 
the doping for $J/t=0.4$ (upper panel) and $J/t=0.2$ (middle panel). The 
results for the variational wave function~(\ref{wf}) (empty symbols) and for 
the FN approximation (filled symbols) are reported. The results for the wave
function without the Jastrow factors (both for spin and density) and magnetic
order parameter are also reported (lower panel).}
\end{figure}

\begin{figure}
\includegraphics[width=0.45\textwidth,clip]{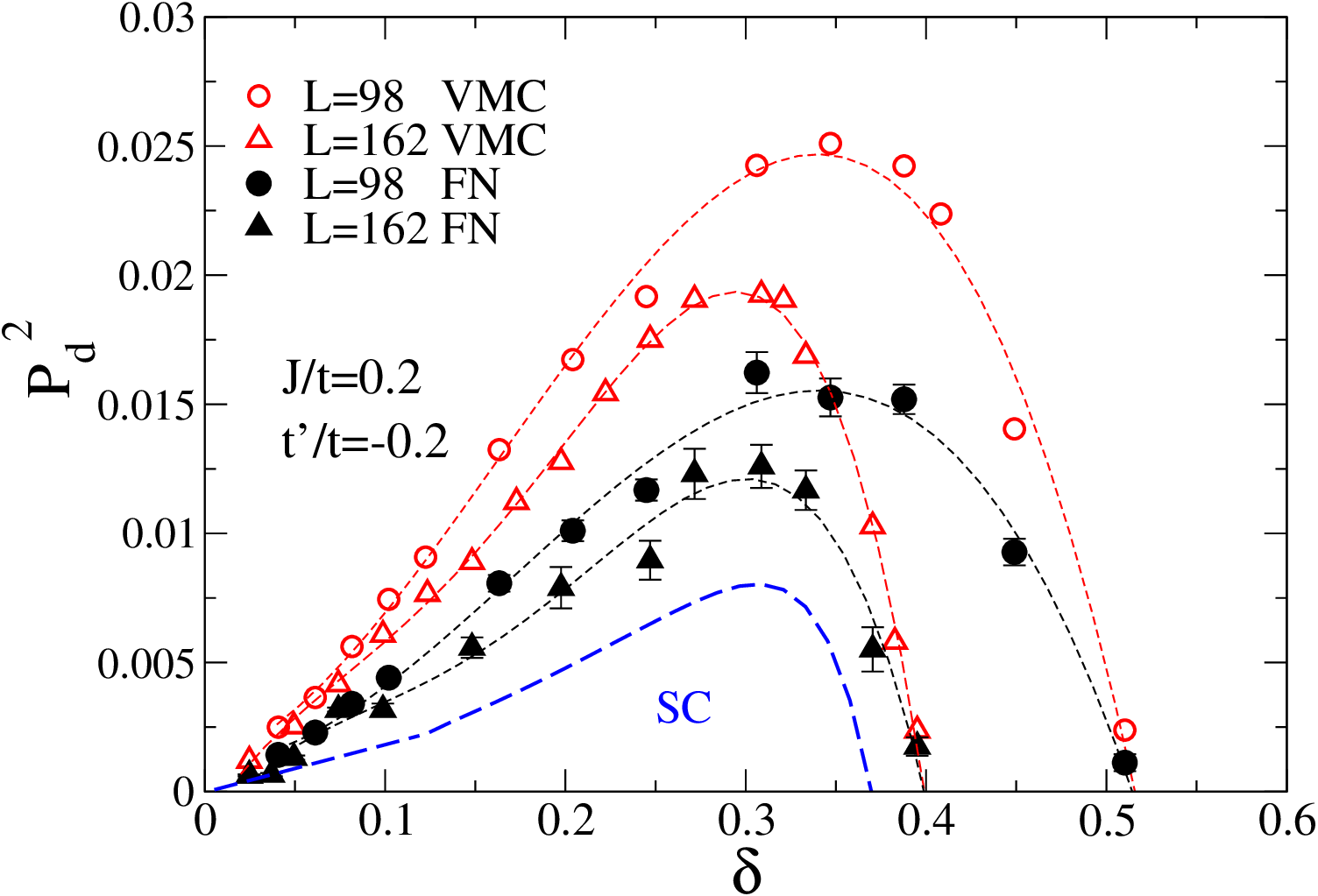}
\caption{\label{fig:pairtprime}
(Color online) The same as in Fig.~\ref{fig:pair} for the $t{-}t^\prime{-}J$ 
model with $J/t=0.2$ and $t^\prime/t=-0.2$. The dashed line indicates 
a tentative estimation for the thermodynamic limit.}
\end{figure}

The superconducting off-diagonal long-range order implies a non-zero value of 
$\langle \Delta^{\mu,\nu}(r) \rangle$ at large distance $r$.
In the following, we consider the pair-pair correlation at the maximum distance
and $\mu=\nu$ (parallel singlets) both for the VMC and the FN approximations
and denote 
$P_d^2=4\lim_{r \to \infty} \langle \Delta^{y,y}(r) \rangle$. 
It is worth noting that, as far as the superconducting correlations are 
concerned, there is no appreciable difference between the results obtained 
with and without the antiferromagnetic order parameter and the long-range 
spin Jastrow factor.
The results for the pure $t{-}J$ model are reported in Fig~\ref{fig:pair}, 
where we report two different values of the antiferromagnetic coupling, 
i.e., $J/t=0.2$ and $J/t=0.4$. In this case, VMC and FN calculations 
are in fairly good agreement, giving a similar superconducting phase diagram. 
Interestingly, the optimal doping, i.e., the doping at which the maximum in the 
pair-pair correlations takes place, occurs in both cases at $\delta \sim 0.2$, 
whereas the actual value of the correlations is proportional to $J/t$. 
At high doping, where antiferromagnetic fluctuations play a minor role, the 
behavior of the pairing is unchanged when $J$ is varied. Although in this 
region there are some size effects, we can safely estimate that 
superconductivity disappears around $\delta \sim 0.35$ and $\delta \sim 0.4$ 
for $J/t=0.2$ and $J/t=0.4$, respectively.

It is worth noting that the density Jastrow term~(\ref{densityjastrow}) is very
important to obtain an accurate estimation of the pairing correlations.
Indeed, the variational results, based on the simple wave function with BCS 
pairing and the on-site Gutzwiller projector (but without the long-range
Jastrow factors) overestimate the pairing correlations at optimal doping by 
at least a factor two. Remarkably, the FN approach is able to correct this 
bias, providing approximately the same results as the one obtained starting 
from the wave function with the long-range Jastrow factor, 
see Fig.~\ref{fig:pair}. This fact shows that the FN method is particularly 
reliable for estimating the pairing correlations. 

\begin{figure}
\includegraphics[width=0.45\textwidth,clip]{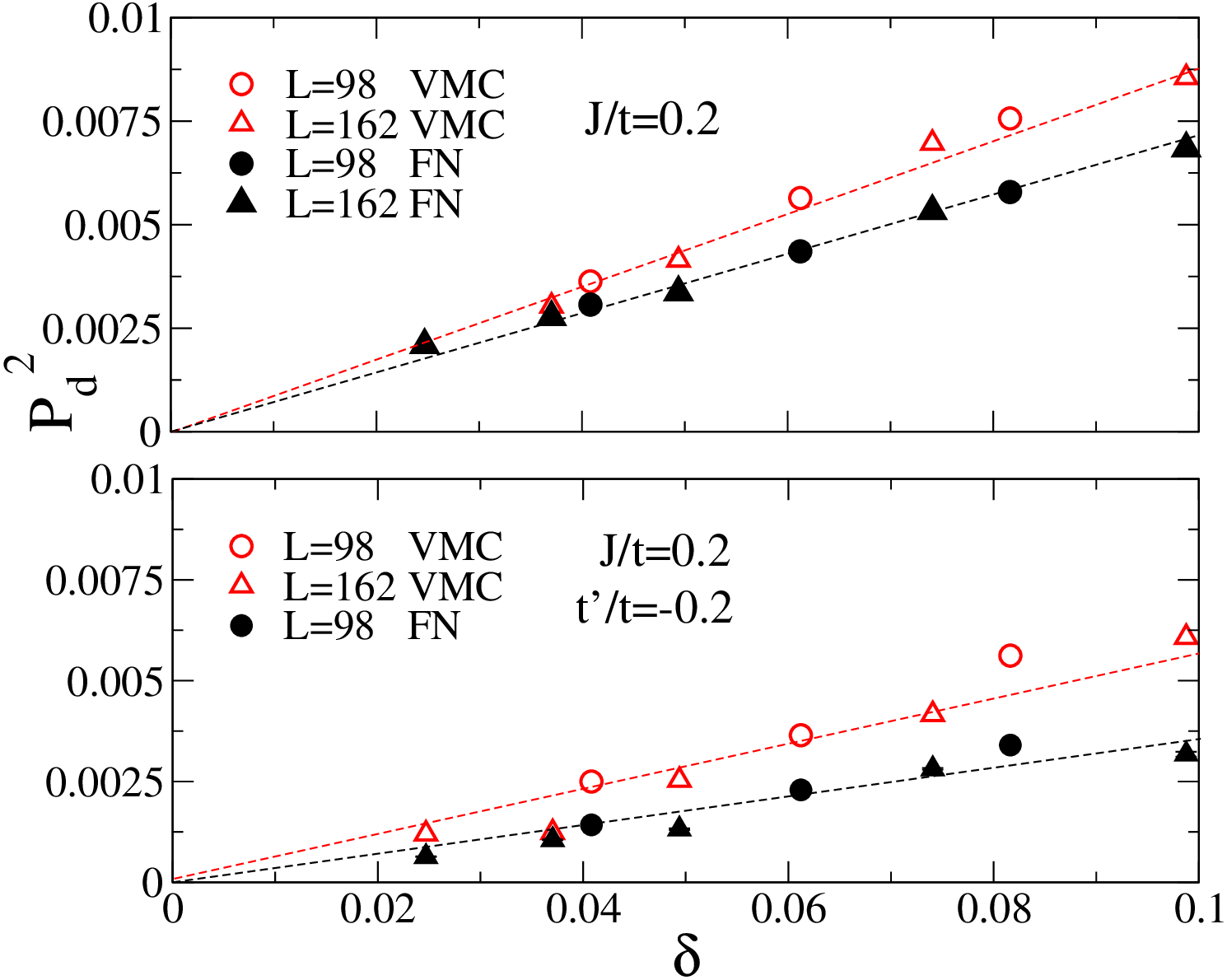}
\caption{\label{fig:pairlow}
(Color online) Detail of the pair-pair correlations reported in 
Figs.~\ref{fig:pair} and~\ref{fig:pairtprime} at low doping.}
\end{figure}

The inclusion of the next-nearest-neighbor hopping induces sizable 
modifications in the pairing correlations, though the qualitative dome-like 
behavior remains unchanged, see Fig.~\ref{fig:pairtprime}. 
At low doping there is a sizable suppression of the superconducting pairing, 
particularly evident after the FN projection, see Fig.~\ref{fig:pairlow}. 
Indeed, while for the pure $t{-}J$ model we clearly obtain a linear
behavior of the pair-pair correlations with $\delta$,~\cite{note} indicating a 
superconducting phase as soon as the Mott insulator is doped, in the case
of a finite $t^\prime$, the FN results could be compatible with a finite
critical doping, below which the system is not superconducting.
This outcome is in agreement with earlier Monte Carlo calculations done by one
of us,~\cite{anisimov} where it was suggested that the extended $t{-}J$ model
with hoppings and super-exchange interactions derived from structural data
of the ${\rm La_2CuO_4}$ compound could explain the main experimental features 
of high-temperature superconducting materials, with a finite critical doping
for the onset of electron pairing. 

\begin{figure}
\includegraphics[width=0.45\textwidth,clip]{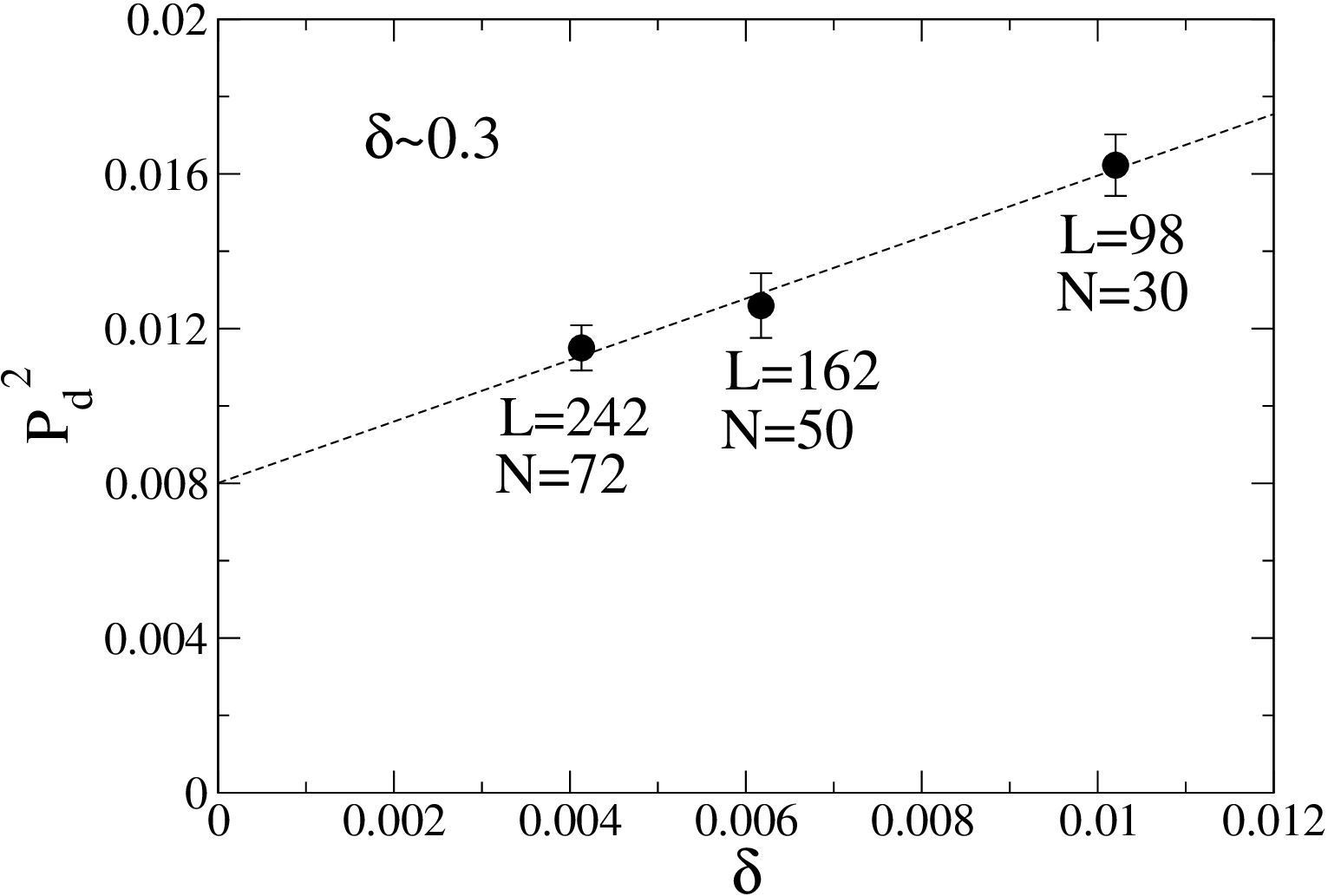}
\caption{\label{fig:scalingpd}
Size scaling of the Pair-pair correlations at the maximum distance for 
$t{-}t^\prime{-}J$ model with $J/t=0.2$ and $t^\prime/t=-0.2$ at 
$\delta \sim 0.3$.}
\end{figure}

Remarkably, from $\delta \sim 0.1$ to $\delta \sim 0.4$ there are huge size 
effects. Though, for $\delta \sim 0.3$, small clusters, e.g., $L=98$, indicate
stronger pairing correlations than the pure $t{-}J$ model without $t^\prime$, 
larger clusters point out a large reduction of $P_d^2$. 
Nonetheless, we have a rather clear evidence that for $\delta \sim 0.3$ there
is a finite superconducting order parameter in the thermodynamic limit,
see Fig.~\ref{fig:scalingpd}.
This strong reduction of the superconducting correlations is a very interesting
effect, demonstrating that the superconducting wave function (even if 
supplemented by magnetic order) deteriorates its accuracy by increasing the 
value of $t^\prime$, that could eventually stabilize competing phases with 
modulation in the charge distribution and/or a magnetic flux through the 
plaquettes.~\cite{oles2} However, for $t^\prime/t=-0.2$, the homogenous 
variational ansatz~(\ref{wf}) provides a lower energy when compared to the one 
used in Ref.~\onlinecite{oles2}.~\cite{capello}
At present, the most accurate FN calculations based on the lowest-energy 
variational ansatz, do not show any tendency towards charge inhomogeneities for
$\delta \lesssim 0.4$. This outcome is important because in principle the FN 
approach can spontaneously induce charge-density wave modulations in the 
ground state, even when the variational wave function before the FN projection 
is translationally invariant.~\cite{beccastripes}
 
\section{Conclusion}\label{conc}

In this paper, we considered the magnetic and superconducting properties of
the $t{-}t^\prime{-}J$ model within VMC and FN approaches.
We showed that for $t^\prime=0$ the ground-state properties can be accurately
reproduced by a state containing both electron pairing and suitable magnetic
correlations, namely a magnetic order parameter in the mean-field 
Hamiltonian that defines the fermionic determinant and a spin Jastrow factor
for describing the spin fluctuations. In this case, we obtain a rather large
magnetic phase, with a critical doping that slightly depends upon the 
super-exchange coupling $J$, i.e., $\delta_c=0.10 \pm 0.01$ and
$\delta_c=0.13 \pm 0.02$ for $J/t=0.2$ and $J/t=0.4$, respectively.
The superconducting correlations show a dome-like behavior and vanish when
the Mott insulator at half filling is approached. Interestingly, compared to 
the RMFT that predicts a quadratic behavior of the pair-pair correlations
as a function of the doping $\delta$, here we found that a linear behavior is
more plausible.

Then, we also reported important modifications due to the presence of a 
finite ratio $t^\prime/t$. The first effect of this further hopping term is
to strongly suppress antiferromagnetic correlations at low doping, shifting
the critical doping to $0.03$ for $t^\prime/t=-0.2$. This is a genuine effect 
of the FN method, since, within the pure variational approach, though the
spin-spin correlations are suppressed with respect to the case of $t^\prime=0$,
the values of the critical doping for these two cases are very similar.
Most importantly, the presence of a finite value of the next-nearest-neighbor
hopping has dramatic effects on the superconducting properties. At small
doping, i.e., $\delta \lesssim 0.1$ there is a sizable suppression of the
electronic pairing, possibly pointing toward a metallic phase in the slightly 
doped regime, as previously suggested by using improved Monte Carlo 
techniques.~\cite{anisimov} Moreover, for $0.1 \lesssim \delta \lesssim 0.4$, 
though small lattices seem to indicate an increasing of superconductivity
compared to the pure $t{-}J$ model, larger clusters show huge size effects 
that strongly renormalize the pairing correlations at large distance. 
However, for the value of $t^\prime$ considered in this work, we are rather
confident that superconducting off-diagonal long-range order takes place in
a considerable hole region. In any case, the huge renormalization of the
electronic pairing for $\delta \sim 0.3$, together with the fact that
the FN results are very different from the VMC ones based on a wave function 
containing pairing (and magnetic order at low doping), suggests the
possibility of the existence of a non-superconducting phase (with magnetic 
fluxes and/or charge order) that could be eventually stabilized by further 
increasing the ratio $t^\prime/t$.  

In conclusion, the main qualitative features of the cuprate phase diagram 
appear rather well reproduced by the $t{-}t^\prime{-}J$ model with a sizable 
next-nearest-neighbor hopping. However, we do not find a sizable enhancement  
of the pairing correlations by increasing the ratio $t^\prime/t$, that looks 
in contradiction with the empirical relation between $t^\prime/t$ and the
value of $T_c$.~\cite{pavarini} However, we have to remark that we only 
considered ground-state properties and we cannot evaluate $T_c$, whose 
relation with the pairing correlations may be highly non trivial, especially 
in a strongly-correlated system, like the $t{-}J$ model. This issue certainly
deserves further work.

We thank useful discussions with S. Bieri, M. Capello, and D. Poilblanc.
This research has been supported by PRIN-COFIN 2005 and CNR-INFM.

\end{document}